\documentstyle[floats,prd,aps]{revtex}
\begin{document}
\draft
\preprint{UM-P-96/74}
\title{The black hole that went away}
\author{ Neil J. Cornish} 
\address{School of Physics, University of Melbourne, Parkville 3052,
Victoria, Australia}
\twocolumn[
\maketitle
\widetext
\begin{abstract}
A purported black hole solution in $(2+1)$-dimensions is shown to be
nothing more than flat space viewed from an accelerated frame.
\end{abstract}
]
\narrowtext

\begin{picture}(0,0)
\put(420,160){{\large UM-P-96/74}}
\end{picture} \vspace*{-0.3 in}

In a recent paper, Kawai~\cite{kawai} describes a remarkable vacuum
solution to Einstein's equations in $(2+1)$-dimensions. The result
is remarkable because it goes against all conventional wisdom. The
claim is that the solution describes a vacuum black hole with a
working Newtonian limit. In contrast, the usual BTZ black hole~\cite{btz}
requires a negative cosmological constant, and $(2+1)$ general relativity
should not have a Newtonian limit~\cite{many}.

The Kawai ``black hole'' has a metric given by~\cite{kawai,k2}
\begin{equation}\label{met}
ds^2=-\left[1+a \ln\left({r \over r_0}\right)\right]^2 dt^2
+\left({r_{0} \over r}\right)^2\left(dr^2 +r^2 d\theta^2\right)\, .
\end{equation}
The coordinates are taken from the conventional range
$-\infty < t <\infty$, $0< r <\infty$, $0\leq \theta \leq 2\pi$ and
$\theta=\theta+2\pi$. The metric appears to
describe a singularity at $r=0$, surrounded by an event horizon at
$r=r_{0}\exp(-1/a)$.

A quick check confirms that (\ref{met}) is indeed a solution to the
vacuum Einstein equations $R_{\mu\nu}=0$. But this is where things start
to unravel. Recall that in
$(2+1)$-dimensions the full Riemann curvature tensor
can always be expressed in terms of the Ricci scalar via the
relation~\cite{deser}
\begin{equation}\label{now}
R^{\mu\nu}_{\;\;\; \kappa\lambda}=\epsilon^{\mu\nu\beta}
\epsilon_{\kappa\lambda\alpha}R^{\alpha}_{\beta} \, .
\end{equation}
Consequently, we know that $R_{\mu\nu\kappa\lambda}=0$ everywhere (there is
no conical singularity at $r=0$). As a result, (\ref{met}) cannot support a
Newtonian limit. 

There is another well known spacetime that is everywhere flat, has an
event horizon, and a coordinate singularity at the origin -- the
Rindler wedge. A simple coordinate transform reveals the Kawai
solution to be nothing more than a Rindler wedge. Transforming
to the cartesian coordinates $x=r_{0}[1+a\ln (r/r_{0})]$, $y=r_{0} \theta$
and rescaling $t$, the metric becomes
\begin{equation}\label{rind}
ds^2=-x^2 dt^2 + dx^2 + dy^2 \, .
\end{equation}
Both $t$ and $x$ have the usual range $(-\infty,\infty)$, while $y$
is topologically compact with period $2\pi r_{0}$. A further set of
standard coordinate transformations reduces the Kawai solution
to Minkowski space. The black hole has gone away.

\end{document}